\documentclass[10pt,journal,a4paper]{IEEEtran}
\makeatletter
\def\ps@headings{%
\def\@oddhead{\mbox{}\scriptsize\rightmark \hfil \thepage}%
\def\@evenhead{\scriptsize\thepage \hfil \leftmark\mbox{}}%
\def\@oddfoot{}%
\def\@evenfoot{}}
\makeatother
\usepackage{multirow}
\usepackage{amssymb}
\usepackage{amsmath}
\usepackage{amsfonts}
\usepackage{graphicx}
\usepackage{subfigure}
\usepackage{subfig}
\usepackage{cite}
\usepackage{enumerate}
\usepackage{url}
\usepackage{mathrsfs}
\usepackage{bm}
\usepackage{clrscode}
\usepackage{subfigure}
\usepackage{amsthm}
\usepackage[ruled]{algorithm2e}
\usepackage{algorithmic}
\usepackage{makecell}
\allowdisplaybreaks[4]
\begin{document}
\title{Large Models for Aerial Edges: An Edge-Cloud Model Evolution and Communication Paradigm}

\author{\IEEEauthorblockN{
		{Shuhang Zhang}, \IEEEmembership{Member, IEEE}, {Qingyu Liu}, \IEEEmembership{Member, IEEE}, {Ke Chen}, \IEEEmembership{Member, IEEE},\\ {Boya Di}, \IEEEmembership{Member, IEEE}, {Hongliang Zhang}, \IEEEmembership{Member, IEEE}, {Wenhan Yang}, \IEEEmembership{Member, IEEE}, \\{Dusit Niyato}, \IEEEmembership{Fellow, IEEE}, {Zhu Han}, \IEEEmembership{Fellow, IEEE}, and
        {H. Vincent Poor}, \IEEEmembership{Life Fellow, IEEE}.
        }\\
\thanks{S. Zhang, K. Chen, and W. Yang are with the Pengcheng Laboratory, Shenzhen, China, 518055. (email: \{zhangshh01, chenk02, yangwh\}@pcl.ac.cn).}
\thanks{Q. Liu is with the School of Electronic and Computer Engineering, Peking University, Shenzhen, China, 518055. (email: qy.liu@pku.edu.cn).}
\thanks{B. Di and H. Zhang are with the Department of Electronics, Peking University, Beijing, China, 100871. (email: \{diboya, hongliang.zhang\}@pku.edu.cn).}
\thanks{D. Niyato is with the College of Computing and Data Science, Nanyang Technological University, Singapore, 639798. (email: dniyato@ntu.edu.sg).}
\thanks{Z. Han is with the Department of Electrical and Computer Engineering at the University of Houston, Houston, TX 77004 USA, and also with the Department of Computer Science and Engineering, Kyung Hee University, Seoul, South Korea, 446-701. (email: hanzhu22@gmail.com).}
\thanks{H.~V.~Poor is with the Department of Electrical Engineering, Princeton University, Princeton, NJ 08544 USA. (email: poor@princeton.edu).}
}
\maketitle

\pagestyle{empty}  
\thispagestyle{empty} 

\begin{abstract}
The future sixth-generation (6G) of wireless networks is expected to surpass its predecessors by offering ubiquitous coverage through integrated air-ground facility deployments in both communication and computing domains. In this network, aerial facilities, such as unmanned aerial vehicles (UAVs), conduct artificial intelligence~(AI) computations based on multi-modal data to support diverse applications including surveillance and environment construction. However, these multi-domain inference and content generation tasks require large AI models, demanding powerful computing capabilities and finely tuned inference models trained on rich datasets, thus posing significant challenges for UAVs. To tackle this problem, we propose an integrated air-ground edge-cloud model evolution framework, where UAVs serve as edge nodes for data collection and small model computation. Through wireless channels, UAVs collaborate with ground cloud servers, providing large model computation and model updating for edge UAVs. With limited wireless communication bandwidth, the proposed framework faces the challenge of information exchange scheduling between the edge UAVs and the cloud server. To tackle this, we present joint task allocation, transmission resource allocation, transmission data quantization design, and edge model update design to enhance the inference accuracy of the integrated air-ground edge-cloud model evolution framework by mean average precision (mAP) maximization. A closed-form lower bound on the mAP of the proposed framework is derived based on the mAP of the edge model and mAP of the cloud model, and the solution to the mAP maximization problem is optimized accordingly. Simulations, based on results from vision-based classification experiments, consistently demonstrate that the mAP of the proposed integrated air-ground edge-cloud model evolution framework outperforms both a centralized cloud model framework and a distributed edge model framework across various communication bandwidths and data sizes.
\end{abstract}

\begin{IEEEkeywords}
Large model, edge intelligence, unmanned aerial vehicle.
\end{IEEEkeywords}

\section{Introduction}\label{Intro}
Integrated air-ground networks are expected to be important components of the sixth generation (6G) of wireless networks, offering seamless connectivity to support a wide array of applications~\cite{MLH2021}. These applications, ranging from surveillance and disaster response~\cite{RZW2020} to environment construction in metaverse~\cite{CCJCLSHWFZ2023}, rely on advanced technologies such as multimodal and generative artificial intelligence~(AI) models~\cite{CZY2023}. These advanced techniques necessitate substantial support from large inference models involving billions of parameters, which is crucial for achieving both high inference accuracy and environmental resilience~\cite{LQCCCH2023}. This requirement, in turn, escalates the demand for ever-increasing computational facility deployments~\cite{RABBBDEFGHH2020}. Consequently, integrated air-ground 6G networks with unmanned aerial vehicles (UAVs) as edge servers, known as one of the six expected use case scenarios of IMT-2030~\cite{IMT2030}, will require ubiquitous services in seamless communications, and encompass the ability to support seamless computing capabilities~\cite{WLKLSXH2020}.

Significant research effort has been devoted to leveraging UAVs as edge computation nodes for various AI applications~\cite{MWL2022}. In~\cite{LCBMRSLK2021}, the authors explored AI modules tailored for UAV-based synthetic aperture radar missions, presenting a comprehensive testbed driven by deep neural networks for object detection. The work in~\cite{YCYQ2020} employed a convolutional neural network deployed on edge UAVs to identify targets within captured video frames, enabling continuous target tracking capabilities. A scalable aerial computing solution applicable for computation tasks of multiple quality levels, corresponding to different computation workloads and computation results of distinct performance was proposed in~\cite{LZCWH2021}, in order to suit the hardware computing capability of edge UAVs. In~\cite{KAAAG2023}, the author proposed a cloud-edge hybrid system architecture, where the edge UAV is responsible for processing AI tasks, and the cloud server is responsible for data storage, manipulation, and visualization.

Despite the promising potential of UAVs as edge AI processors, the frameworks in~\cite{LCBMRSLK2021,YCYQ2020,LZCWH2021,KAAAG2023} are incapable of supporting UAVs working as edge AI nodes in envisioned 6G networks in two aspects. \emph{First}, the onboard computing capacity of UAVs is insufficient for the demanding applications expected for 6G networks. Previous studies~\cite{LCBMRSLK2021,YCYQ2020,LZCWH2021,KAAAG2023} show that UAVs can only perform inference tasks that require low computing capability, driven by models with a few million parameters, such as YOLOv7~\cite{WBL2023}. However, 6G networks are expected to support applications like disaster response and environmental construction, which demand multimodal large models with billions of parameters~\cite{BAM2018}, such as SORA~\cite{SORA} and Gemini~\cite{G2023}. \emph{Second}, the computing capabilities of edge UAVs are insufficient for model training, and thus the edge models cannot be updated onboard~\cite{XNZKXMH2023}, which results in limited robustness and accuracy for edge AI services~\cite{GGALMCCRR2020}. Consequently, the accuracy of onboard inference models degrades severely with environmental variations~\cite{YCXYGLQH2021}. For the above reasons, a new framework that supports cooperations between edge UAVs and ground cloud servers with powerful computing capabilities is needed in order to provide large model driven data processing services for edge UAVs~\cite{TFLH2021,ZZS2020}.

To tackle the above problems, in this paper, we propose a new integrated air-ground edge-cloud model evolution framework based on a joint data and model communication paradigm. In the proposed framework, each edge UAV is responsible for collecting sensory data, with the flexibility to conduct local computing using an onboard small model or upload the data to the cloud server for large model analysis. The uploaded data contains extracted feature data, together with partial residual mapping data~\cite{DLYHG2020}, which can be dynamically adjusted according to the communication data rate between the UAV and a cloud server~\cite{MZJZWW2020}. To improve the performance of the edge model, the cloud server also transmits model updating information to the edge UAV. We formulate an integrated air-ground model cooperation optimization problem to enhance the inference accuracy performance of the entire network. The design of the formulated problem encompasses task allocation between the edge UAV and the cloud server, along with considerations for the overhead of feature transmission, residual mapping data transmission, and model update transmission so as to maximize the mean average precision~(mAP) of the UAV and the cloud server jointly.

Note that several problems and challenges warrant careful consideration in the design of this integrated air-ground edge-cloud model evolution framework. \emph{First}, it is important to define a performance metric for the proposed framework. This metric will serve as a crucial foundation for optimizing task allocation and communication resource allocation between the edge UAV and the cloud server. \emph{Second}, the uplink data transmission facilitates cloud model computation with high mAP, while the downlink model updates enhance the mAP of the edge model. Therefore, with limited wireless communication bandwidth, a thorough investigation of the trade-off between uplink and downlink resource allocation is essential. \emph{Third}, given the constraints of limited uplink transmission bandwidth, the UAV faces the decision of transmitting either low-resolution feature data for more tasks or high-resolution residual mapping data for fewer tasks to the cloud server. An in-depth analysis of the trade-off between feature transmission and residual mapping data transmission is thus important.

By addressing the aforementioned challenges, our contributions are summarized as follows:
\begin{enumerate}
\item \textbf{Framework Proposal:} We introduce a new integrated air-ground edge-cloud model evolution framework, facilitating the handling of cloud models for edge UAVs' data and supports the evolution of edge models on UAVs with assistance from a ground server. This framework accommodates three distinct data transmission streams: the feature stream, data stream, and model stream. The amount of data transmitted on each stream can be dynamically adjusted in accordance with the communication bandwidth of the wireless network.
\item \textbf{Problem Formulation and Analysis:} Building upon the proposed framework, we formulate the joint edge-cloud mAP maximization problem, which involves optimizing edge-cloud task allocation, uplink-downlink bandwidth allocation, residual mapping data quantization design, and model update overhead design. To address the formulated problem, we derive an expression for the joint edge-cloud mAP as a function of the edge model mAP and cloud model mAP, and optimize the formulated problem under arbitrary transmission bandwidth constraints based on the derived formula.
\item \textbf{Performance Evaluation:} The proposed framework's performance is evaluated using results from vision-based classification experiments. Simulation results demonstrate the mAP gain achieved by our framework when compared to centralized and distributed computing frameworks across different wireless transmission parameters and data sizes. It is concluded that the edge model handles the majority of tasks with small communication bandwidth and large data size, with most bandwidths allocated to small model updating. Conversely, the cloud model handles the majority of tasks with large communication bandwidth and small data size, with most bandwidths allocated to data uploading.
\end{enumerate}

The rest of this paper is organized as follows. In Section~\ref{Scheme}, we propose our integrated air-ground edge-cloud model evolution framework in detail. Section~\ref{System Model Sec} outlines the system model of the integrated air-ground edge-cloud model evolution framework with one edge UAV and one cloud server. In Section~\ref{Problem}, the joint edge-cloud mAP maximization problem is formulated, and the resulting mixed integer programming problem is decomposed into two subproblems. In Section~\ref{SolutionSec}, we solve the mAP maximization problem, and analyze the properties of the integrated air-ground edge-cloud model evolution framework. Simulation results are presented in Section~\ref{Simulations}. Finally, the conclusions are drawn in Section~\ref{Conclusions}. The abbreviations and notations used in this paper are listed in Tables~\ref{Abbreviation} and~\ref{Notation}, respectively.

\begin{table}[!tpb]
\centering
\caption{Abbreviations}\label{Abbreviation}
\begin{tabular}{|c|c|}
\hline
\textbf{Abbreviation} & \textbf{Full Name}\\
\hline
6G & Sixth Generation\\
\hline
UAV & Unmanned Aerial Vehicle\\
\hline
AI & Artificial Intelligence \\
\hline
mAP & Mean Average Precision\\
\hline
OTA & Over the Air\\
\hline
BS & Base Station\\
\hline
NR & New Radio\\
\hline
LTE & Long Term Evolution\\
\hline
IoU & Intersection over Union\\
\hline
PRC & Precision-Recall Curve\\
\hline
TP & True Positive\\
\hline
FP & False Positive\\
\hline
FN & False Negative\\
\hline
\end{tabular}
\end{table}

\begin{table}[!tpb]
\centering
\caption{Notation}\label{Notation}
\begin{tabular}{|c|c|}
\hline
\textbf{Notation} & \textbf{Meaning}\\
\hline
 $N$ & Number of frames generated per second\\
\hline
$x$ & Pixels per frame\\
\hline
$\beta$ & Task allocation ratio\\
\hline
$\Psi$ & Set of frames analysed at the cloud\\
\hline
$\Phi$ & Set of frames with residual mapping data transmission\\
\hline
$\bar{F}$ & Average size of extracted feature of a frame\\
\hline
$R_F$ & Transmission data rate of the feature stream\\
\hline
$\rho$ & \makecell[c]{Fraction of frames in $\Psi$with \\residual mapping data transmission}\\
\hline
$\bar{\bm{b}}$ & Residual mapping data quantization bit \\
\hline
$R_D$ & Transmission data rate of the data stream\\
\hline
$B_u$ & Uplink transmission bandwidth\\
\hline
$B_d$ & Downlink transmission bandwidth\\
\hline
$M$ & Model update overhead\\
\hline
$S_u$ & Spectrum efficiency of uplink transmission\\
\hline
$S_d$ & Spectrum efficiency of downlink transmission\\
\hline
$mAP$ & \makecell[c]{mAP of the integrated air-ground \\edge-cloud model evolution framework}\\
\hline
$mAP_L$ & mAP of the cloud model\\
\hline
$mAP_S$ & mAP of the edge model\\
\hline
$r_L^k$ & Recall value of the cloud model with the $k$th IoU\\
\hline
$r_S^k$ & Recall value of the edge model with the $k$th IoU\\
\hline
$p_L^k$ & Precision value of the cloud model with the $k$th IoU\\
\hline
$p_S^k$ & Precision value of the edge model with the $k$th IoU\\
\hline
\end{tabular}
\end{table}

\section{Integrated Air-Ground Edge-Cloud Model Evolution Framework}\label{Scheme}

\begin{figure*}[!tpb]
\centering
\includegraphics[width=4.5in]{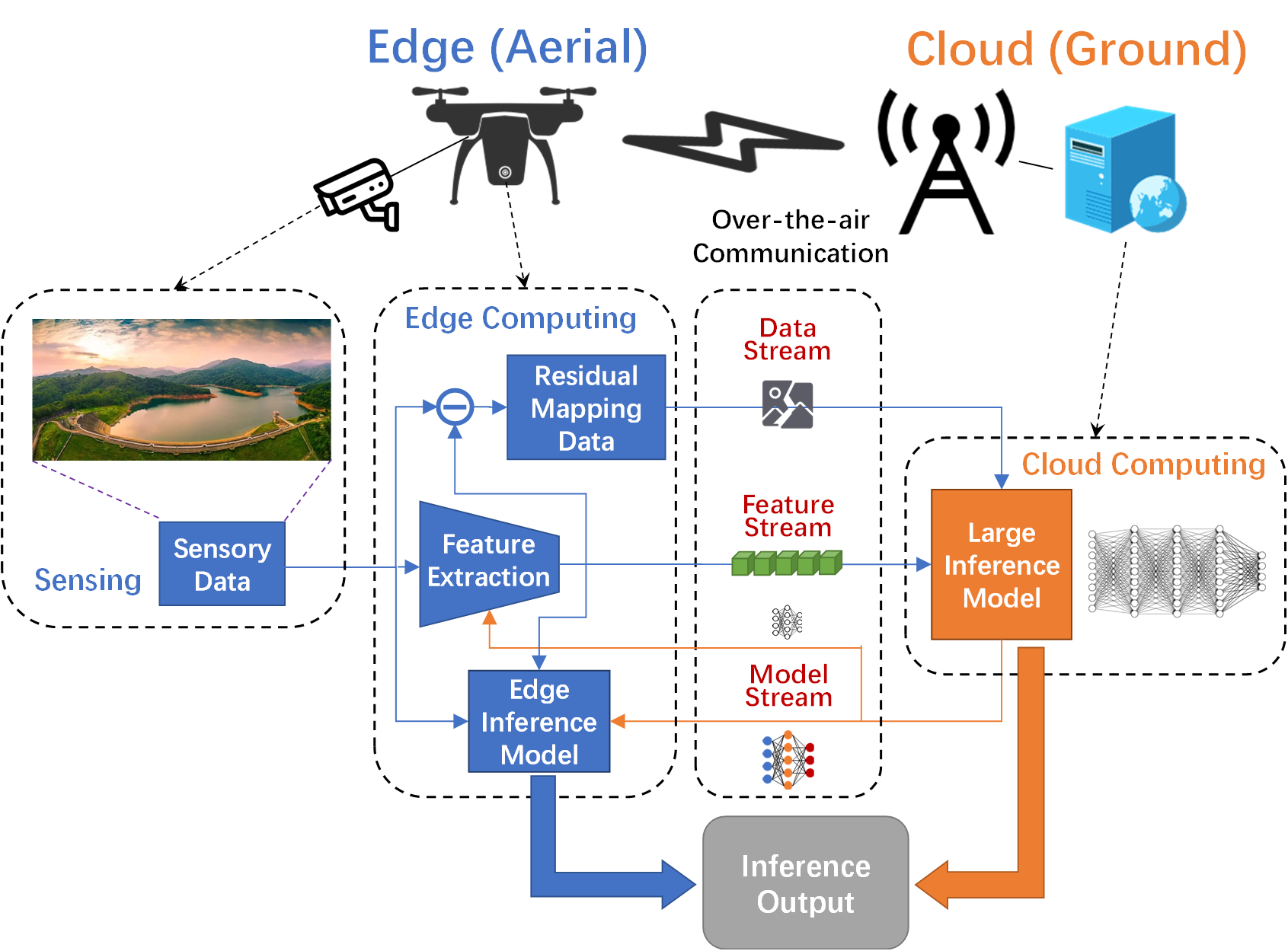}
\caption{Paradigm for an integrated air-ground edge-cloud model evolution framework.}
\label{Framework}
\end{figure*}

In this section, we introduce our integrated air-ground edge-cloud model evolution framework that facilitates simultaneous edge computing and cloud computing. The proposed integrated air-ground edge-cloud model evolution framework is illustrated in Fig.~\ref{Framework}, which consists of edge nodes (i.e., UAVs) and cloud nodes (i.e., cloud servers). For clear illustration, only one edge UAV and one cloud node is presented. The front-end UAV, equipped with an onboard data collector (e.g., a video camera) and edge computing module, serves as a remote sensor and edge server. The back-end ground cloud server functions as a central node for enhanced analysis and recognition. Given the inherent instability of wireless communication bandwidth, the integrated air-ground edge-cloud model evolution framework requires a flexible communication paradigm design. This includes dynamic task allocation, on-demand residual mapping data transmission, and flexible edge model updates.

To be specific, each edge UAV is responsible for collecting data for subsequent data analysis and recognition. For simplicity, we take a vision data classification task as an example. The analysis and recognition on each frame is referred to as a \emph{task}. The tasks can be either executed at the UAV with an onboard edge model or at the ground cloud server with a cloud model. To facilitate cloud model analysis at the cloud server, the edge UAV first extracts sensory data using an onboard feature extraction model, and then transmits the extracted features of visual data to the cloud server via over-the-air~(OTA) transmission, known as the feature stream. For further enhancement of inference performance at the cloud server, supplemental data providing detailed visual descriptions beyond the information extracted by the feature model, referred to as residual mapping data, can be transmitted to the cloud server over idle OTA transmission resources with adjustable resolution, known as the data stream. In response to tasks and data from various domains, the feature extraction and edge inference models at the UAV are upgraded by receiving model updating data from the cloud server with flexible overhead, known as the model stream.

Recently, several supportive works have been studied for implementing the above three streams in the integrated air-ground edge-cloud model evolution framework~\cite{DZSLHP2020}. For feature stream, the compact feature representation technique ensures high-efficiency feature extraction and data compression, leading to a reduced overhead of the feature stream to a few Kbps~\cite{DLYHG2020}. For residual mapping data in the data stream, an intelligent coding technique facilitates the efficient representation of image/video, enabling dynamic encoding of the video stream into a practical and suitable level~\cite{MZJZWW2020}. For edge model updates, a model compression and incremental updating technique allows for dynamic model update via the model stream, thus providing in-time response to the task and data from various domains~\cite{CBBNC2022}.

The aforementioned studies have demonstrated the viability of incorporating the feature stream, data stream, and model stream within the integrated air-ground edge-cloud model evolution framework. However, there still remains an issue in the investigation of OTA communications. To be specific, it is necessary to study the two following initial aspects. \emph{First}, considering that the overhead, i.e., the size of OTA transmitted data, of the three streams can be dynamically adjusted, it is important to study the function of the performance metric of each stream with respect to the communication data rate. \emph{Second}, the optimization of resource allocation for wireless transmissions of the three streams should be investigated jointly to maximize system performance within the constraints of limited transmission bandwidth. These solutions to the above challenges are the foundation of implementing our integrated air-ground edge-cloud model evolution framework, and require in-depth study. With the supportive techniques, the proposed integrated air-ground edge-cloud model evolution framework can significantly expand the applications of cloud model supported edge AI across various scenarios, such as precision agriculture, target searching, and disaster area rescue, harnessing the capabilities of AI techniques~\cite{BRR2022,FFKTCDHAC2023}.


\section{System Model}\label{System Model Sec}
\begin{figure}[!tpb]
\centering
\includegraphics[width=3.5in]{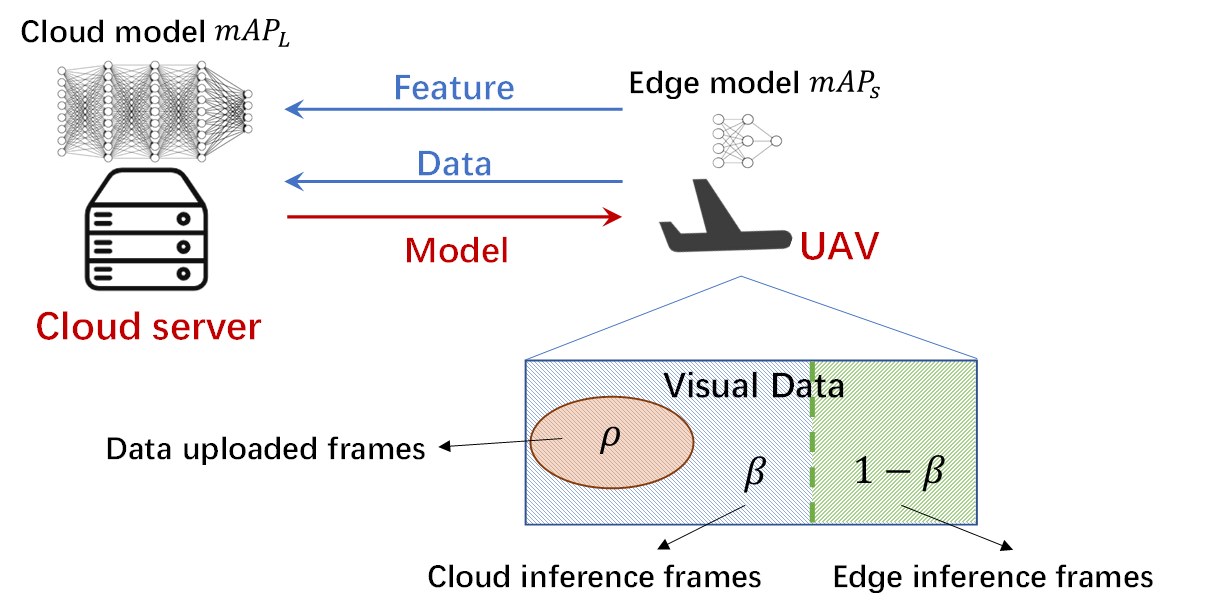}
\caption{System model for an integrated air-ground edge-cloud model evolution framework.}
\label{model}
\end{figure}

In this section, we introduce a fundamental system model of the integrated air-ground edge-cloud model evolution framework. The system model contains one ground cloud server and one front-end UAV as a joint sensing, edge computing, and communication node, as shown in Fig.~\ref{model}. Note that the solution to the design of this scenario also works on each of the UAVs in the multi-UAV scenario. Due to space limitation, the part of the design on multi-UAV scenario, such as the resource allocation among different UAVs, will be studied in our future works. The UAV is equipped with an onboard camera and is tasked with capturing visual data, subsequently collaborating with the cloud server to perform target classification, in order to support various edge services, such as disaster response and geographic identification.

Due to the constraints in energy and computing capabilities, the UAV faces challenges in independently performing high-accuracy visual classification with the edge model onboard. The cloud server can provide assistance to the UAV in target classification through cloud model computation and edge model updates. The cloud server is connected to a ground base station (BS), capable of communicating with the UAV via OTA transmission networks such as new radio (NR) and long-term evolution (LTE). As introduced in Section~\ref{Scheme}, three streams are transmitted via OTA data transmissions, namely model stream, feature stream, and data stream. The model stream is a downlink transmission, while feature stream and data stream are uplink transmissions.

We assume that the onboard camera of the UAV captures frames at a frequency of $N$ per second, where each frame contains $x$ pixels, and each pixel is quantized into $b$ bits. The value of $N$, $x$ and $b$ are determined by the hardware of the onboard camera and the energy consumption constraint of the UAV. Consequently, the visual data generation rate of the UAV is $D=N\cdot x\cdot b$. As depicted in Fig.~\ref{model}, a fraction $\beta$ number of frames is uploaded to the cloud server via OTA transmissions for visual target classification, while the remaining fraction $1-\beta$ number of the frames is processed at the UAV with the edge model. We denote the set of frames analysed at the cloud server by $\Psi$, with $|\Psi|=\beta N$. The features of the frames in $\Psi$ are extracted at the UAV for subsequent processing at the cloud server. Let $\bar{F}$ be the average size of the extracted feature of a frame, and the transmission data rate of the feature stream is given by
\begin{equation}\label{Feature Stream Rate}
R_{F}=\bar{F}\beta N.
\end{equation}

For enhanced classification accuracy at the cloud server, the residual mapping data of a fraction $\rho$ of frames can be transmitted from the UAV to the cloud server. We denote the set of frames whose residual mapping data is sent to the cloud server by $\Phi$, with $|\Phi|=\rho N$. As the residual mapping data needs to be combined with the extracted feature at the cloud server for image reconstruction, the set $\Phi$ only contains the frames analysed at the cloud server, i.e., $\Phi\subseteq\Psi$. With limited OTA transmission bandwidth, it is necessary to properly quantize the residual mapping data. Let $\bm{b}=\{\hat{b}_i\}, \forall i \in \Phi$ be the quantization parameter of the residual mapping data, where $\hat{b}_i$ represents the number of quantization bits per pixel of frames $i$, which is selected from a set of discrete values, denoted by $\Omega$. The transmission data rate of the data stream can be expressed as
\begin{equation}\label{Data Stream Rate}
R_{D}=\sum_{i\in \Phi}x\hat{b}_i, \forall \hat{b}_i \in \Omega.
\end{equation}

The sum of the transmission data rate of the feature stream and data stream should be no larger than the upload capacity of the UAV, i.e.,
\begin{equation}\label{UL Rate}
R_{F}+R_{D}\leq B_u\cdot S_u,
\end{equation}
where $B_u$ is the bandwidth for UAV uplink transmission, and $S_u$ is the spectrum efficiency of UAV uplink transmission. The spectrum efficiency can be considered as available value with proper channel measurement techniques as studied in~\cite{JZWZDWZ2020,YXQCC2021}, regardless of the wireless propagation environment.

The cloud server accumulates a substantial volume of feature and residual mapping data from edge UAV,\footnote{In this paper, we analyze a single UAV scenario. The same paradigm can be extended to a network with multiple UAVs by properly measuring the spectrum efficiency of each UAV.} and subsequently updates the model for edge UAV to enhance its inference accuracy. The model update information is then transmitted to the UAV with an average overhead of $M$ bits per second, representing the data rate of the model stream. Importantly, the data rate of the model stream cannot exceed the download capacity of the UAV, i.e.,
\begin{equation}\label{DL Rate}
M\leq B_d\cdot S_d,
\end{equation}
where $B_d$ is the bandwidth for UAV downlink transmission, and $S_d$ is the spectrum efficiency of UAV downlink transmission. It is also assumed that the total bandwidth allocated for UAV-BS communication is $B$, which satisfies
\begin{equation}\label{Bandwidth}
B_u+B_d\leq B.
\end{equation}

In the study of this paper, the spectrum efficiencies of the uplink and downlink transmissions, i.e., $S_u$ and $S_d$, are considered as constants with arbitrary values. The impact factors on $S_u$ and $S_d$, such as the UAV trajectory, transmission beamforming, transmission power design, and interference management can be considered as independent designs from this work. Works on optimizing the spectrum efficiency of the integrated air-ground edge-cloud model evolution framework will be studied in future works.

\section{Problem Formulation and Decomposition}\label{Problem}
In this section, we first formulate the mAP maximization problem for the integrated air-ground edge-cloud model evolution framework described in Section~\ref{System Model Sec}, and then decompose the problem into two subproblems for further analysis.

\subsection{mAP Maximization Problem Formulation}\label{ProblemFormulation}
The mAP of the integrated air-ground edge-cloud model evolution framework is determined by three factors: the mAP of the cloud server large model $mAP_L$, the mAP of the UAV edge small model $mAP_S$, and the fraction of target classification frames analysed at the cloud server $\beta$. For simplicity, we denote the function representing mAP as $mAP=f(mAP_L, mAP_S, \beta)$. The expression and properties of the function $f(\cdot)$ will be studied in Section~\ref{SolutionSec}.

We assume that the model at the cloud server is well trained with stable performance, and the variable $mAP_L$ is determined by the quality of the feature and residual mapping data transmitted from the UAV~\cite{RMSSTLK2017}. As studied in~\cite{YHHDL2021}, the mAP of the feature based inference converges to a stable level with an overhead much smaller than the residual mapping data size. Therefore, we consider that a feature extraction method with fixed overhead is adopted for each of the frames in set $\Psi$. The mAP of the cloud model $mAP_L$ is a function of the proportion of residual mapping data transmission $\rho$ and the corresponding quantization bits $\hat{b}_i$, denoted by $mAP_L=g(\rho, \hat{b}_i), \forall i\in \Phi$ for simplicity. The expression of function $g(\cdot)$ may vary for different tasks and models, and can be fitted with experimental data from related studies, such as~\cite{WWWY2021}.

For the edge UAV, it is capable of obtaining lossless data of all frames. The mAP performance is affected by the inference accuracy of the edge model, which is determined by the model update provided by the cloud server. The mAP of the edge model $mAP_S$ is expressed as $mAP_S=h(M), M_{min}\leq M\leq M_{max}$, where $M_{min}$ and $M_{max}$ are the minimum and maximum model update overhead, respectively. The value of $M_{min}$ and $M_{max}$ are related to parameters such as model update algorithm, UAV computing capability and power consumption constraints. The expression of function $h(\cdot)$ may vary for different tasks, and can be fitted with experimental data of related studies, such as~\cite{CBBNC2022}.

To maximize the mAP of the integrated air-ground edge-cloud model evolution framework, it is essential to jointly optimize the edge-cloud task allocation, uplink-downlink bandwidth allocation, residual mapping data transmission design, and model update overhead design. The problem can be formulated as
\begin{subequations}\label{formulate problem}
\begin{align}
\max_{\beta, \rho, \bm{b}, B_d, B_u, M} \ & mAP=f(mAP_L, mAP_S, \beta),\label{ObjFunc}\\
s.t. & mAP_L=g(\rho, \hat{b}_i), \forall i\in \Phi,\label{Cons2}\\
& 0\leq \rho\leq \beta\leq 1,\label{Cons3}\\
& \hat{b}_i \in \Omega, \forall i \in \Phi,\label{Cons4}\\
& mAP_S=h(M),\label{Cons5}\\
& M_{min}\leq M\leq M_{max},\label{Cons6}\\
& R_{F}+R_{D}\leq B_u\cdot S_u,\label{Cons7}\\
& M\leq B_d\cdot S_d,\label{Cons8}\\
& B_u+B_d\leq B.\label{Cons9}
\end{align}
\end{subequations}

Objective function~(\ref{ObjFunc}) represents the maximization of the mAP for the integrated air-ground edge-cloud model evolution framework, which is a function of variables $mAP_L, mAP_S,$ and $\beta$. Constraint~(\ref{Cons2}) captures the mAP function of the cloud model. Constraint~(\ref{Cons3}) specifies that the fraction of frames with residual mapping data transmission should not exceed the fraction of frames analysed at the cloud server. Constraint~(\ref{Cons4}) pertains to the quantization constraint for residual mapping data transmission. Constraint~(\ref{Cons5}) represents the mAP function of the edge model at the UAV, and constraint~(\ref{Cons6}) imposes the constraint on the overhead of the edge model update. Finally, constraints~(\ref{Cons7})-(\ref{Cons9}) involve the transmission data size constraints for the feature stream, data stream, and model stream, respectively.

Problem (\ref{formulate problem}) poses significant challenges for direct solution due to two primary reasons. \emph{First}, it is a mixed-integer programming problem that encompasses both discrete variables in $\bm{b}$ and continuous variables $\beta, \rho, B_d, B_u, M$, which is NP hard. \emph{Second}, the convexity of this problem cannot be ensured, as the convexity of the experimentally fitted functions $g(\cdot)$ and $h(\cdot)$ remains uncertain. In the subsequent analysis, we aim to decompose problem (\ref{formulate problem}) into two subproblems: the data stream design subproblem, and the feature/model stream design subproblem, and analyse the two subproblems in sequence. With such problem decomposition, the discrete variables in $\bm{b}$ can be separated from parameters $\beta, B_d, B_u, M$ to simplify the complicated formulated problem in~(\ref{formulate problem}), and discussions on the convexity of $g(\cdot)$ and $h(\cdot)$ can be decoupled into two independent subproblems.

\subsection{Problem Decomposition}

\subsubsection{Data Stream Design Subproblem}
In the data stream design subproblem, our attention is directed towards the uplink data stream, which influences the mAP of the cloud model inference at the server. This includes the design of the set of frames for residual mapping data transmission $\rho$, and the quantization bits of the residual mapping data of each transmitted frame $\bm{b}$. The parameters associated with edge model update, task allocation, and transmission resource allocation are treated as fixed values and are not subject to optimization in this subproblem. The first subproblem is formulated to maximize the mAP of the cloud model, by optimizing the proportion of frames with residual mapping data transmission, and their corresponding quantization bits. The first subproblem can be formulated as follows:
\begin{subequations}\label{subproblem1}
\begin{align}
\max_{\rho, \bm{b}} \ & mAP_L,\label{Sub1ObjFunc}\\
s.t. & mAP_L=g(\rho, \hat{b}_i), \forall i\in \Phi,\label{Sub1Cons1}\\
& 0\leq \rho \leq \beta,\label{Sub1Cons2}\\
& \hat{b}_i \in \Omega, \forall i \in \Phi,\label{Sub1Cons3}\\
& R_{F}+R_{D}\leq B_u\cdot S_u.\label{Sub1Cons4}
\end{align}
\end{subequations}
Constraints~(\ref{Sub1Cons1})-(\ref{Sub1Cons4}) are related to the data stream, which have been introduced in Section~\ref{ProblemFormulation}.

\subsubsection{Feature/Model Stream Design Subproblem}
Assuming that the parameters related to the data stream have been optimized from the solution of subproblem (\ref{subproblem1}), our attention in this subproblem is devoted to designing the parameters associated with the feature stream and model stream. The second subproblem is formulated to maximize the joint mAP of the cloud model and edge model. This is achieved by optimizing the fraction of target classification frames allocated to the edge UAV and the cloud server, the transmission bandwidth allocated to the uplink and downlink transmissions, and the overhead of the edge model update. The second subproblem can be formulated as below,
\begin{subequations}\label{subproblem2}
\begin{align}
\max_{\beta, B_d, B_u, M} \ & mAP=f(mAP_L, mAP_S, \beta),\label{Sub2ObjFunc}\\
s.t.& 0\leq \rho\leq \beta\leq 1,\label{Sub2Cons2}\\
& mAP_S=h(M),\label{Sub2Cons3}\\
& M_{min}\leq M\leq M_{max},\label{Sub2Cons4}\\
& R_{F}+R_{D}\leq B_u\cdot S_u,\label{Sub2Cons5}\\
& M\leq B_d\cdot S_d,\label{Sub2Cons6}\\
& B_u+B_d\leq B.\label{Sub2Cons7}
\end{align}
\end{subequations}
Constraints~(\ref{Sub2Cons2})-(\ref{Sub2Cons7}) are related to joint cloud-edge computing, which have been introduced in Section~\ref{ProblemFormulation}.

\section{Solution and Analysis for Integrated Air-Ground Edge-Cloud Model Evolution Framework}\label{SolutionSec}
In this section, we focus on solving the main mAP maximization problem in~(\ref{formulate problem}). The two subproblems~(\ref{subproblem1}) and~(\ref{subproblem2}) are solved in Sections~\ref{Feature and data} and~\ref{Uplink and downlink}, respectively. An overall solution to problem~(\ref{formulate problem}) and the subsequent analysis of the solution are presented in Section~\ref{Analysis}.

\subsection{Solution to Data Stream Design Subproblem}\label{Feature and data}
In this subsection, we focus on the design to data stream, and solve subproblem~(\ref{subproblem1}). The parameters related to feature stream and model stream are considered to be fixed. Since the quantization bits of each frame can be different, the mAP of different frames may vary. Denote the mAP of the cloud model on analysing frame $i$ by $mAP_L^i$.\footnote{The mAP of a frame serves as a metric to measure the accuracy of an inference model, distinct from the statistical definition of mAP defined in~(\ref{AppendixAmAPEqu}).} In order to solve subproblem~(\ref{subproblem1}), we first explain important properties and assumptions related to function $g(\cdot)$ in constraint~(\ref{Sub1Cons1}).

\textbf{Remark 1:} The mAP of frame $i$, i.e. $mAP_L^i$, monotonically increases with respect to the quantization bits of its residual mapping data $\hat{b}_i$.

\textbf{Assumption 1:} The mAP of frame $i$, i.e. $mAP_L^i$, is a concave function of $\hat{b}_i,$ considered as a continuous variable with $\hat{b}_i \in \Omega$.

\textbf{Assumption 2:} The mAP of frame $i$, i.e. $mAP_L^i$, is a concave function of $\hat{b}_i,$ considered as a continuous variable with $\hat{b}_i \in \Omega\cup \{0\}$, where $\hat{b}_i=0$ corresponds to the case with no residual mapping data transmission.

\emph{Remark 1} emphasizes that precise residual mapping data contributes to improved mAP performance at the cloud server, which is intuitively understandable. \emph{Assumption 1} is derived from observations across various experiments on multiple datasets~\cite{WWWY2021,LSZYL2020,ALHT2020}. Although it lacks theoretical proof, it holds true for most current studies. Therefore, \emph{Assumption 1} is considered valid for the majority of existing visual-based classification tasks. \emph{Assumption 2} is an extended statement of \emph{Assumption 1} that covers the case where the residual mapping data of a frame is not transmitted to the cloud server, with the quantization bit being 0. In this case, only extracted features are sent to the cloud server as the input of the cloud model.

However, it is important to note that \emph{Remark 1} and \emph{Assumption 1} do not ensure the convexity of subproblem~(\ref{subproblem1}), since $mAP_L^i$ and $mAP_L$ are not equivalent. In the following, we further provide two theorems related to $mAP_L$, providing a basis for solving subproblem~(\ref{subproblem1}).

\textbf{Theorem 1:} Without the discrete quantization bits constraint~(\ref{Sub1Cons3}), the solution that maximizes $mAP_L$ satisfies $\hat{b}_1=\hat{b}_2=\cdots =\hat{b}_i, \forall i\in \Phi$.

\begin{proof}
See Appendix A.
\end{proof}

\textbf{Theorem 2:} When \emph{Assumption 2} is satisfied, the residual mapping data of all the frames in $\Psi$ should be sent to the cloud server with the same quantization bits, i.e., $\hat{b}_1=\hat{b}_2=\cdots =\hat{b}_i, \forall i\in \Psi$.
\begin{proof}
See Appendix B.
\end{proof}

With Theorems 1 and 2, subproblem~(\ref{subproblem1}) can be solved as follows. Variables $R_{F}$, $B_u$ and $S_u$ in (\ref{subproblem1}) are given, and the constraint~(\ref{Sub1Cons4}) can be converted to $\sum_{i\in \Phi}x\hat{b}_i\leq B_u\cdot S_u-R_{F}$. When \emph{Assumption 2} is satisfied, we first set $\rho=\beta$ and $\hat{b}_1^{opt}=\cdots =\hat{b}_i^{opt}=\frac{B_u\cdot S_u-R_{F}}{|\Psi|}, \forall i\in \Psi$. If the value of $\hat{b}_1^{opt}$ does not satisfy constraint~(\ref{Sub1Cons3}), the value of elements in $\bm{b}$ are selected from $\hat{b}^{l}$ and $\hat{b}^{u}$, where $\hat{b}^{l}$ and $\hat{b}^{u}$ are the two closest value to $\hat{b}_1^{opt}$, satisfying $\hat{b}^{l}<\hat{b}_1^{opt}<\hat{b}^{u}$ and $\hat{b}^{u}, \hat{b}^{l} \in \Omega\cup \{0\}$. A ratio of $\lceil \frac{\hat{b}^{opt}-\hat{b}^{l}}{\hat{b}^{u}-\hat{b}^{l}}\rfloor$ frames are quantized with $\hat{b}^{u}$ bits for the residual mapping data, while a ratio of $\lceil\frac{\hat{b}^{u}-\hat{b}^{opt}}{\hat{b}^{u}-\hat{b}^{l}}\rfloor$ frames are quantized with $\hat{b}^{l}$ bits for the residual mapping data, where $\lceil\cdot\rfloor$ is the function for obtaining the closest integer.

In cases where \emph{Assumption 2} is not met, we propose a heuristic-based method to address subproblem~(\ref{subproblem1}). This heuristic approach involves calculating the maximum data rate for residual mapping data transmission, denoted as $B_u\cdot S_u-R_{F}$. We introduce the concept of mAP increment efficiency for each frame, which represents the increase in mAP with unit increment in the data quantization bits. The strategy then prioritizes the allocation of the remaining communication resources to frames with the highest mAP increment efficiency. This iterative process continues until all communication resources are effectively allocated.

\subsection{Solution to Feature/Model Stream Design Subproblem}\label{Uplink and downlink}
As we have solved the design to data stream related parameters in the last subsection, in this part, we aim to optimize the parameters related to feature stream and model stream to solve subproblem~(\ref{subproblem2}). To facilitate this, we introduce a theorem that outlines the properties of the joint mAP involving both the cloud model and the edge model.

\textbf{Theorem 3:} The joint mAP of cloud model and edge model is a function of recall-precision pairs\footnote{The definition of recall-precision pair has been explained in Appendix A for the proof of \emph{Theorem 1}.} of the cloud model and the edge model, which can be expressed as
\begin{equation}\label{Theorem3}
\begin{split}
mAP=\frac{1}{2}\cdot \sum_{k=1}^K \left(\frac{1}{\frac{\beta}{r_L^k}+\frac{1-\beta}{r_S^k}}-\frac{1}{\frac{\beta}{r_L^{k-1}}+\frac{1-\beta}{r_S^{k-1}}}\right)\\ \times \left(\frac{1}{\frac{\beta}{p_L^k}+\frac{1-\beta}{p_S^k}}+\frac{1}{\frac{\beta}{p_L^{k-1}}+\frac{1-\beta}{p_S^{k-1}}}\right),
\end{split}
\end{equation}
where $r_L^k$ and $p_L^k$ are the recall and precision values of the cloud model with the $k$th intersection over union (IoU), and $r_S^k$ and $p_S^k$ are the recall and precision values of the edge model with the $k$th IoU, respectively.
\begin{proof}
See Appendix C.
\end{proof}

\emph{Theorem 3} proves the relation between $mAP$ and the precise-recall values. However, the optimization variables in~(\ref{subproblem2}) are not directly related to the precise-recall values. In the subsequent discussion, we delve deeper into the relationship among $mAP$, $mAP_L$, and $mAP_S$ established upon the insights provided by \emph{Theorem 3}.

\textbf{Theorem 4:} The joint mAP of cloud model and edge model satisfies
\begin{equation}
mAP\geq \frac{mAP_L\cdot mAP_S}{(1-\beta)mAP_L+\beta mAP_S}.
\end{equation}
\begin{proof}
See Appendix D.
\end{proof}

In \emph{Theorem 4}, we derive the lower bound of $mAP$ as a function of $mAP_S$, $mAP_L$, and $\beta$. To deepen our understanding, our goal is to establish a closed-form relationship between $mAP$ and $mAP_S$, $mAP_L$, and $\beta$ under specific conditions. The subsequent \emph{Theorem 5} focuses on scenarios where the $mAP$ performances of the cloud model and the edge model are of the same magnitude, which is a common case for most of the tasks and inference models in related studies~\cite{WWWY2021,LSZYL2020,ALHT2020}, and a closed-form expression of $mAP$ is illustrated.

\textbf{Theorem 5:} When the constraint $r_L^k-r_S^k\ll r_L^k, p_L^k-p_S^k\ll p_L^k, \forall 1\leq k\leq K$ is satisfied, the joint mAP of the cloud model and edge model can be approximated as
\begin{equation}\label{mAPApproximate}
mAP\approx \frac{mAP_L\cdot mAP_S}{(1-\beta)mAP_L+\beta mAP_S}.
\end{equation}
\begin{proof}
As proved in Appendix D, the variable $\zeta^k=p^k r^k$ can be converted to
\begin{equation}
\zeta^k=\frac{\zeta_L^k \zeta_S^k}{(1-\beta)\zeta_L^k+\beta \zeta_S^k-\beta(1-\beta)\Delta},
\end{equation}
with $\Delta=(p_L^k - p_S^k)\cdot (r_L^k - r_S^k)$. When constraints $r_L^k-r_S^k\ll r_L^k$, and $p_L^k-p_S^k\ll p_L^k$ are satisfied, we have $\Delta\ll \zeta_L^k$, and thus
\begin{equation}\label{zetaEqu}
\zeta^k\simeq\frac{\zeta_L^k \zeta_S^k}{(1-\beta)\zeta_L^k+\beta \zeta_S^k}.
\end{equation}
Since $mAP$ is a linear combination of a series of $\zeta^k$, the relationship in~(\ref{zetaEqu}) also holds for the $mAP$, and equation~(\ref{mAPApproximate}) holds.
\end{proof}

Even in cases where the constraint $r_L^k-r_S^k\ll r_L^k, p_L^k-p_S^k\ll p_L^k, \forall 1\leq k\leq K$ is not strictly met, (\ref{mAPApproximate}) can still be considered as an lower bound of $mAP$, to solve the optimization problem~(\ref{subproblem2}). The convexity of equation (\ref{mAPApproximate}) with respect to $mAP_L$ and $mAP_S$ can be obtained by calculating its Hessian matrix, i.e.,
\begin{equation}\label{Hessian}
\begin{split}
H
=&
\left[
\begin{array}{cc}
\frac{\partial ^2 (mAP)}{\partial (mAP_L)^2} & \frac{\partial ^2 (mAP)}{\partial (mAP_L) \partial (mAP_S)}  \\
\frac{\partial ^2 (mAP)}{\partial (mAP_S) \partial (mAP_L)} & \frac{\partial ^2 (mAP)}{\partial (mAP_S)^2}
\end{array}
\right]\\
=&\frac{2\beta(1-\beta)}{((1-\beta)mAP_L+\beta mAP_S)^3}\times\\
&\left[
\begin{array}{cc}
-(mAP_S)^2 & mAP_L mAP_S  \\
mAP_L mAP_S & -(mAP_L)^2
\end{array}
\right].
\end{split}
\end{equation}
As shown in (\ref{Hessian}), the first-order and second-order principal minor of the Hessian matrix are both non-positive. Therefore, equation (\ref{mAPApproximate}) is a concave function with respect to $mAP_L$ and $mAP_S$.

After analysing the convexity of $f(\cdot)$, we study the convexity of $mAP_L$ with respect to $R_{F}+R_{D}$, to examine the convexity of subproblem~(\ref{subproblem2}). As studied in Section~\ref{Feature and data}, $mAP_L$ is a concave function with respect to $R_{D}$. Since the value of $R_{F}$ dost not affect $mAP_L$, $mAP_L$ can be considered as a concave function with respect to $R_{F}+R_{D}$. Given that the size of uplink transmitted data $R_{F}+R_{D}$ is a linear function of the uplink transmission bandwidth $B_u$, $mAP_L$ is also a concave function with respect to $B_u$.

According to \emph{Theorem 5}, it can be observed that the expression of $mAP$ is concave with respect to $mAP_L$ and $mAP_S$, when $r_L^k-r_S^k\ll r_L^k$, and $p_L^k-p_S^k\ll p_L^k, \forall 1\leq k\leq K$ are satisfied. Moreover, the function $h(\cdot)$ in constraint~(\ref{Sub2Cons3}) has been fitted to be concave in existing studies~\cite{CBBNC2022}. Under these conditions, subproblem~(\ref{subproblem2}) is a concave function with respect to variables $\beta, B_d, B_u, M$, and can be addressed using convex optimization methods. Even when $r_L^k-r_S^k\ll r_L^k, p_L^k-p_S^k\ll p_L^k, \forall 1\leq k\leq K$ is not satisfied, a lower bound solution can be obtained by approximating function $f(\cdot)$ following~(\ref{mAPApproximate}).

\subsection{Overall Algorithm and Analysis}\label{Analysis}
In this part, we first summarize the overall algorithm for solving the integrated air-ground edge-cloud model evolution framework design problem~(\ref{formulate problem}), and then analyse the impact factors on the solution to this problem.

The approach to solve the mAP maximization problem~(\ref{formulate problem}) is outlined in Algorithm~\ref{Solution}. Initially, we derive an optimal value of $mAP_L$ concerning $B_u$ by solving subproblem~(\ref{subproblem1}). Subsequently, we tackle subproblem~(\ref{subproblem2}) to determine the solution for variables $\beta, B_d, B_u$, and $M$. The solution of $B_u$ is then substituted into subproblem~(\ref{subproblem1}), yielding the final solution for $\beta$ and $\rho$. When the conditions $r_L^k-r_S^k\ll r_L^k$, and $p_L^k-p_S^k\ll p_L^k, \forall 1\leq k\leq K$ hold true, \emph{Theorem 5} is applicable, and the optimal solution can be obtained. Alternatively, if the conditions are not satisfied, a suboptimal solution is obtained considering $mAP= \frac{mAP_L\cdot mAP_S}{(1-\beta)mAP_L+\beta mAP_S}$. According to \emph{Theorem 4}, the true value of $mAP$ is no less than $\frac{mAP_L\cdot mAP_S}{(1-\beta)mAP_L+\beta mAP_S}$, and the solution obtained by the proposed algorithm serves as a lower bound for~(\ref{formulate problem}).

\begin{algorithm}[!t]
\caption{Joint cloud model and edge model design for the integrated air-ground edge-cloud model evolution framework.}
\begin{algorithmic}[1]\label{Solution}
\STATE {\textbf{Input:} Variables $B,S_d,\Omega,M_{min},M_{max},N$, functions $g(\cdot)$, $h(\cdot)$;}
\STATE {Solve subproblem~(\ref{subproblem1}) to obtain the function of maximized $mAP_L$ with respect to $B_u$;}
\STATE {\textbf{If} $r_L^k-r_S^k\ll r_L^k, p_L^k-p_S^k\ll p_L^k, \forall 1\leq k\leq K$;}
\STATE {\quad Solve concave optimization problem~(\ref{subproblem2}) to obtain optimal values of $\beta, B_d, B_u, M$;}
\STATE {\textbf{Else}}
\STATE {\quad Obtain a lower bound of $mAP$ with a sub-optimal solution of $\beta, B_d, B_u, M$;}
\STATE {\textbf{EndIf}}
\STATE {Solve for $\bm{b}, \rho$ corresponds to the $B_u$ obtained in problem~(\ref{subproblem2});}
\STATE {\textbf{Output:} Task allocation variable $\beta$, data quantization variables $\rho, \bm{b}$, communication variables $B_d, B_u$, Model update variable $M$;}
\end{algorithmic}
\end{algorithm}

\textbf{Theorem 6:} The complexity of the proposed Algorithm~\ref{Solution} is $O(N\cdot B^2)$.

\begin{proof}
As shown in Algorithm~\ref{Solution}, subproblem~(\ref{subproblem1}) and~(\ref{subproblem2}) are solved sequentially with different values of $B_u$. The enumerations of $B_u$ is in proportion to the bandwidth $B$. In each enumeration, subproblem~(\ref{subproblem1}) is first solved as introduced in Section~\ref{Feature and data}. When \emph{Assumption 2} is satisfied, variables $\bm{b}, \rho$ can be solved with a complexity of $O(N)$. When \emph{Assumption 2} is not satisfied, the heuristic-based method allocates bandwidth resources to each frame sequentially, with a complexity of $O(N\cdot B)$. Therefore, the complexity of the solution to subproblem~(\ref{subproblem1}) is $O(N\cdot B)$. As introduced in Section~\ref{Uplink and downlink}, subproblem~(\ref{subproblem2}) can be solved with convex optimization method directly. Since the optimization variables $\beta, B_d, B_u, M$ are all elements rather than vectors, the complexity of the optimization is a constant $C$. In summary, the total complexity of the proposed Algorithm~\ref{Solution} is $B\cdot O(N\cdot B+C)=O(N\cdot B^2)$.
\end{proof}

After solving the formulated problem in~(\ref{formulate problem}), we analyse the relation between the optimal design for $mAP_L$ and $mAP_S$ with respect to the wireless communication capacity.

As analysed in the above subsections, the mAP of the integrated air-ground edge-cloud model evolution framework is determined by the mAP at the cloud server, the mAP at the edge UAV, and the fraction of target classification frames analysed at the cloud server. Given an unit of transmission bandwidth, the mAP of the integrated air-ground edge-cloud model evolution framework can be improved in three options:
\begin{enumerate}[(1)]
\item Enhancing the overhead for model update $M$ to improve $mAP_S$;
\item Enhancing the quantization bits of the frames $\bm{b}$ in set $\Phi$ to improve $mAP_L$;
\item Enhancing the fraction of frames analysed at the cloud server $\beta$ to improve the mAP of the framework.
\end{enumerate}

For the optimal communication paradigm, the changing rate of $mAP$ with the above three options should be equal to a unit communication bandwidth variation. Otherwise, a better solution can be obtained by reducing the transmission bandwidth allocated to the option with lower mAP changing rate, while improving that of the option with higher changing rate. In what follows, a theorem that examines the relationship between the mAP of the cloud model at the server and the mAP of the edge model at the UAV under corresponding communication and computing configurations is given. This analysis provides a quantitative understanding of the trade-off between improving the model at the edge node and achieving high mAP performance at the cloud server.

\textbf{Theorem 7:} When \emph{Theorem 2} and \emph{Theorem 5} hold, the relation between $mAP_L$ and $mAP_S$ for the optimal data-model communication paradigm can be expressed as
\begin{equation}\label{Theorem5Equation}
\begin{split}
&\frac{mAP_L}{mAP_S} =
\sqrt{(\frac{N(\bar{F}+\hat{b}_i)mAP_L}{mAP_L-mAP_S}-\frac{mAP_S}{\frac{\emph{d} (h(M))}{\emph{d} M}}\cdot\frac{S_u}{S_d})} \\&\times \sqrt{\frac{\partial g(\rho, \hat{b}_i)}{\partial \hat{b}_i}\cdot \frac{1}{|\Psi|}}.
\end{split}
\end{equation}

\begin{proof}
See Appendix E.
\end{proof}

From \emph{Theorem 7}, we conclude the impact factors of the optimal design to the capacity of the classification models at the edge UAV and the cloud server as below.

\textbf{Theorem 8:} The relation between the capacity of the classification models at the edge UAV and the cloud server is determined by the following factors:
\begin{enumerate}[(1)]
\item The wireless transmission quality ($S_u$ and $S_d$);
\item The feature extraction and data quantization condition for frames ($\bar{F}$ and $\hat{b}_i$);
\item The function characteristics of the frame quantization and model update ($g(\cdot)$ and $h(\cdot)$);
\item The number of frames to analyse at the cloud server ($N$ and $\Psi$);
\end{enumerate}

The closed-form function of the above parameters to the mAPs of the edge model and cloud model is provided, which offers comprehensive guidance for training the edge model in networks with varying communication and computing capabilities. This also shows that the mAP performance gain of the integrated air-ground edge-cloud model evolution framework is at the cost of high OTA bandwidth/spectrum efficiency for feature stream, data stream, and model stream overhead transmissions, and highly compressed algorithms for feature extraction and model update.

\section{Simulation Results}\label{Simulations}
In this section, we evaluate the performance of the proposed integrated air-ground edge-cloud model evolution framework with joint task allocation, transmission resource allocation, transmission data quantization optimizations, and edge model update design. For comparison, we compare the proposed framework with three baseline frameworks: a centralized cloud model framework, a distributed edge model framework, and exhaustive search for the integrated air-ground edge-cloud model evolution framework.

\begin{enumerate}
\item \emph{Centralized cloud model framework:} In this framework, the edge UAV has no classification capability. It transmits the extracted features and quantized residual mapping data of all the frames to the cloud server for target classifications. The quantization bits of the frames is determined by the bandwidth and spectrum efficiency of the OTA uplink transmission.
\item \emph{Distributed edge model framework:} In this framework, the edge UAV performs local classifications. The cloud server only transmits model update for the edge UAV according to the OTA transmission capability.
\item \emph{Exhaustive search:} In this framework, the proposed integrated air-ground edge-cloud model evolution framework is adopted. The task allocation, transmission resource allocation, transmission data quantization optimizations, and edge model update design are selected by enumerating over $10^8$ candidate variable combinations for mAP maximization. The performance can be considered as an upper bound of the proposed framework.
\end{enumerate}

In this simulation, we take a visual-based classification task as an example, the value of the related parameters are presented in Table~\ref{Simulation Parameter}. The experimental data is based on a classification task on CIFAR10 dataset~\cite{CIFAR10}. The model at the edge UAV is a ResNet18 with 11.7M parameters, and the model at the cloud server is set as a ResNet 101 with 45M parameters~\cite{HZRS2016}. The training process and model updating process at the edge UAV follows the methods proposed in~\cite{CDWLHWG2019}. The function $g(\cdot)$, representing $mAP_L$ with respect to the quantization bits, is fitted as results from experiments with the proposed algorithm in~\cite{WWWY2021}. Similarly, the function $h(\cdot)$, representing $mAP_S$ with respect to the model update overhead, is fitted using results from experiments with the proposed algorithm in~\cite{CDWLHWG2019}. Note that the proposed system and its corresponding optimizations can be applied to various tasks with different models and datasets.

\begin{table}[!tpb]
\centering
\caption{Simulation Parameters}\label{Simulation Parameter}
\begin{tabular}{|c|c|}
\hline
\textbf{Parameter} & \textbf{Value}\\
\hline
Number of sensing frames generated per second $N$ & 10\\
\hline
Number of pixels per frame $x$ & $10^7$\\
\hline
Average data size of the extracted feature $\bar{F}$ & 0.86kbps \\
\hline
OTA bandwidth $B$ & 10 MHz\\
\hline
Uplink spectrum efficiency $S_u$ & 2.55 bit/s/Hz\\
\hline
Downlink spectrum efficiency $S_d$ & 5 bit/s/Hz\\
\hline
Maximum model update overhead $M_{max}$ & 230 kbps\\
\hline
Minimum model update overhead $M_{min}$ & 23 Mbps\\
\hline
\end{tabular}
\end{table}

\begin{figure}[!tpb]
\centering
\includegraphics[width=2.5in]{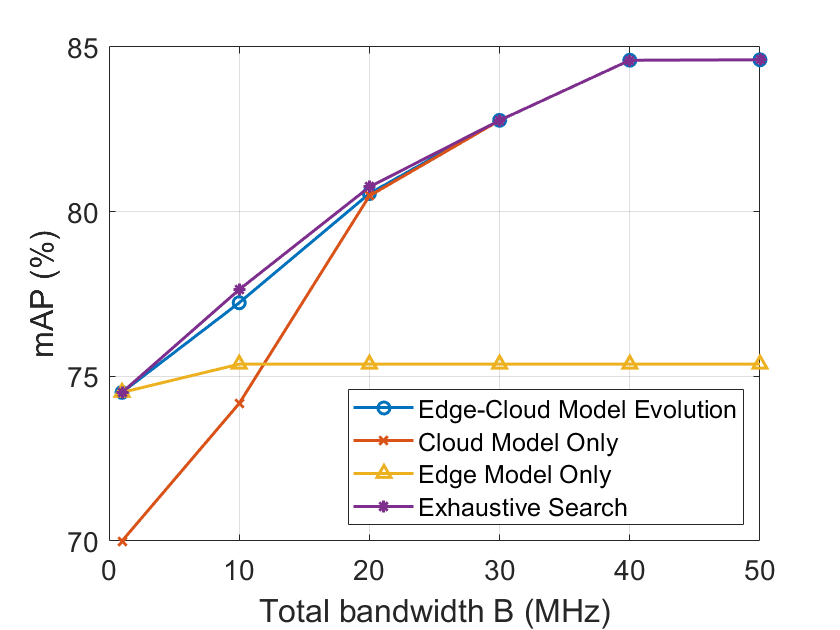}
\caption{Total bandwidth $B$ versus mAP.}
\label{Simulation1}
\end{figure}

In Fig.~\ref{Simulation1}, the mAP of the integrated air-ground edge-cloud model evolution framework is illustrated with different total transmission bandwidths. It is shown that the mAP increases with larger transmission bandwidth, and converges to a stable value when the total bandwidth is over 40 MHz. The convergent mAP value implies that with sufficiently large bandwidth, all the frames can be uploaded to the cloud server for high mAP analysis. The performance of the proposed framework is comparable to that of the edge model framework when the bandwidth is less than 5 MHz, where most tasks are performed at the edge model due to limited data transmission capability. When the bandwidth is larger than 20 MHz, the mAP of the proposed framework and the cloud model framework are close, with most of the residual mapping data sent to the cloud server for high mAP analysis. As a dynamic combination of cloud model computation and edge model computation, the proposed edge-cloud model evolution framework always outperforms the cloud model framework and the edge model framework with different values of $B$ by dynamically adjusting the communication resources to obtain the optimal $mAP_L$ and $mAP_S$. The mAP performance gap between the proposed method and the exhaustive search is within 0-0.5\% in all cases.

\begin{figure}[!tpb]
\centering
\includegraphics[width=2.5in]{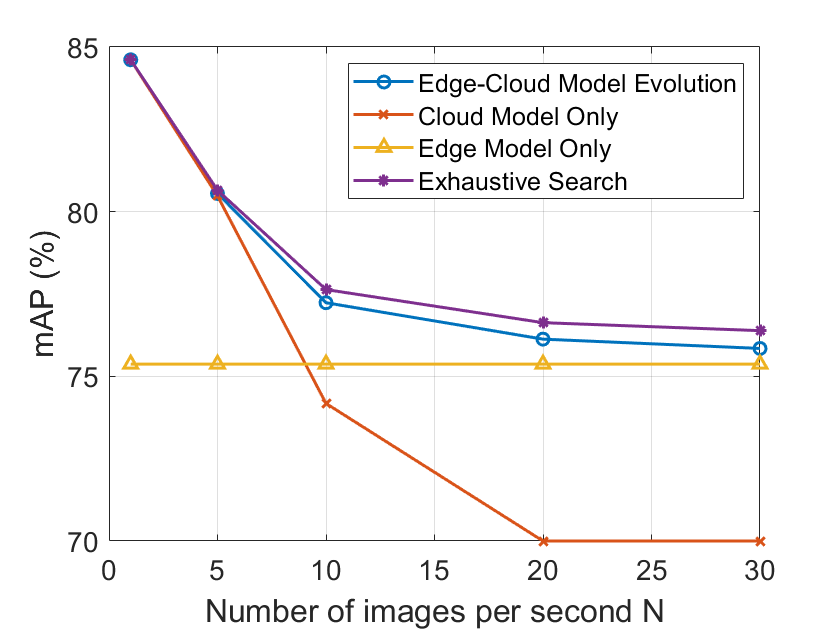}
\caption{Number of frames per second $N$ versus mAP.}
\label{Simulation2}
\end{figure}

In Fig.~\ref{Simulation2}, we evaluate the mAP with different number of frames generated per second. Given fixed communication bandwidth, a larger number of generated frames corresponds to less average residual mapping data transmission for each frame, thereby leading to mAP decrement. In the case of the edge model framework, the mAP is only determined by the model update, which is independent from the number of frames generated per second, and the mAP is a constant value with different $N$. For the proposed edge-cloud model evolution framework, although the mAP reduces with a larger value of $N$, the mAP is lower bounded by that of the edge model. The mAP performance of the proposed framework consistently outperforms both the cloud model framework and the edge model framework across different values of $N$, and the performance gap to the mAP of the exhaustive search is always less than 0.65\%.

\begin{figure}[!tpb]
\centering
\includegraphics[width=2.5in]{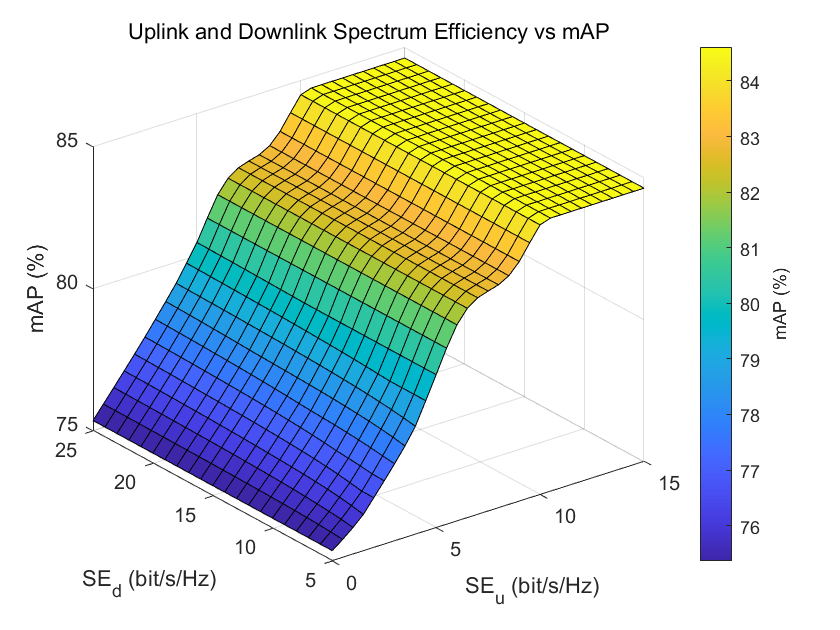}
\caption{Uplink and downlink spectrum efficiency versus mAP.}
\label{Simulation3}
\end{figure}

\begin{figure}[!tpb]
\centering
\includegraphics[width=3.5in]{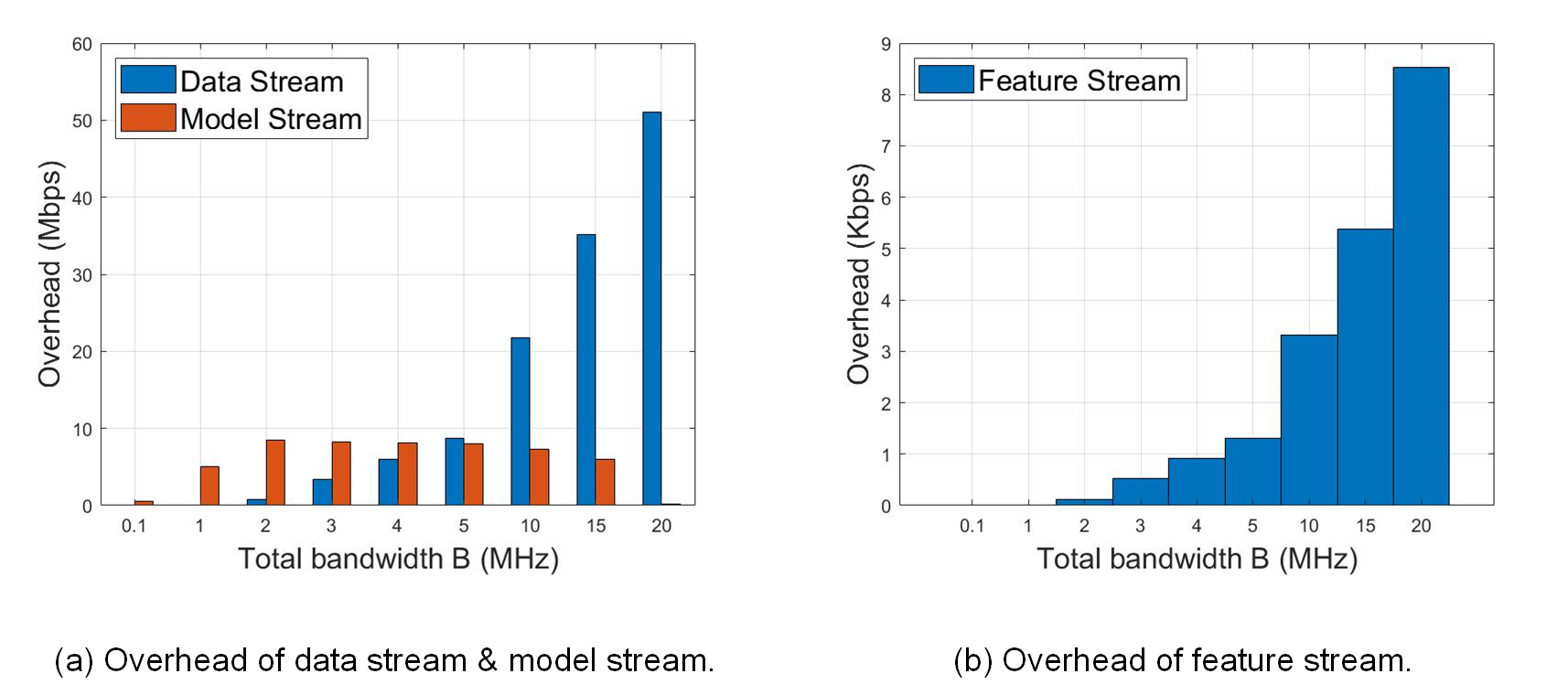}
\caption{Total bandwidth $B$ versus overhead of each stream.}
\label{Simulation4}
\end{figure}

Fig.~\ref{Simulation3} demonstrates the impact of the spectrum efficiency of uplink and downlink transmissions on the mAP of the integrated air-ground edge-cloud model evolution framework. Both uplink spectrum efficiency $S_u$ and downlink spectrum efficiency $S_d$ exhibit a positive correlation with $mAP$. The influence of $S_u$ on $mAP$ is more significant than that of $S_d$, due to the larger overhead of uplink residual mapping data compared to the downlink model update, as further illustrated
in Fig.~\ref{Simulation4}. When $S_u$ exceeds 10 bit/s/Hz, $mAP$ is no longer affected by $S_d$. This is because, under this condition, all frames are sent to the cloud server for high-accuracy mAP analysis, rendering the update of the edge model unnecessary.

In Fig.~\ref{Simulation4}, we present the trade-off between enhancing the edge model and uploading data to the cloud model. The overhead of the data stream, model stream, and feature stream in the proposed framework are illustrated under different total transmission bandwidths. The overhead of the feature stream is considerably lower than that of the data stream and the model stream, facilitating cloud model computation with a low uplink transmission bandwidth. As depicted in Fig.~\ref{Simulation4} (a), when the total bandwidth is less than 2 MHz, the model stream dominates the OTA transmission overhead. This suggests that with a relatively low communication capacity, most of the tasks are performed at the edge model, emphasizing the significance of $mAP_S$ over $mAP_L$ under this condition. As the bandwidth exceeds 2 MHz, the overhead of the data stream increases significantly, and becomes much larger than that of the model stream when the bandwidth surpasses 10 MHz. This indicates that, with increased bandwidth, the majority of target classification tasks are handled by the cloud server. The overhead of model streams starts decreasing when the bandwidth exceeds 5 MHz, as the fraction of target classification frames analysed at the cloud server diminishes. Consequently, the impact of $mAP_S$ on $mAP$ becomes less pronounced than that of $mAP_L$. Additionally, Fig.~\ref{Simulation4} (b) suggests that the overhead of the feature stream steadily increases with the total bandwidth $B$ due to the transmission of features from a larger number of frames.

\begin{figure}[!tpb]
\centering
\includegraphics[width=2.5in]{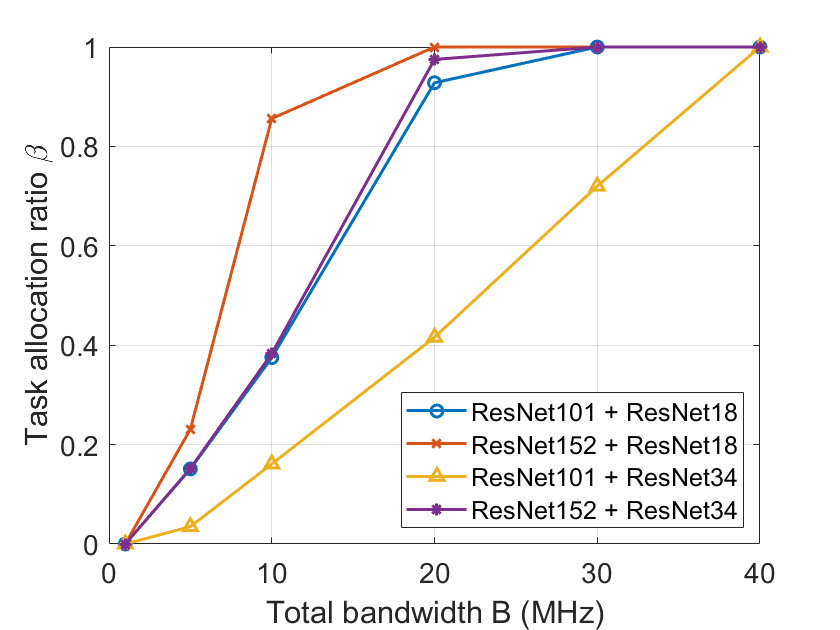}
\caption{Total bandwidth $B$ versus task allocation ratio $\beta$.}
\label{Simulation5}
\end{figure}

In Fig.~\ref{Simulation5}, we investigate the impact of total bandwidth on fraction of target classification frames analysed at the cloud server, i.e., $\beta$. As analysed in Section~\ref{SolutionSec}, the optimal solution to $\beta$ is affected by the mAP at the cloud server and the edge UAV. Therefore, we study four different cases with various model configurations. For the purpose of this analysis, we assume that with improved computing capabilities, both the edge UAV and the cloud server can upgrade their models to larger architectures: a ResNet34 model with 20M parameters for the UAV and a ResNet101 model with 110M parameters for the cloud server~\cite{HZRS2016}.\footnote{The enhanced models can be seen as a representation of future capabilities with stronger computing power and improved inference models. The specific mAP percentage increment can be adjusted dynamically with specific models.} The results indicate that a higher classification accuracy at the cloud server corresponds to a larger value of $\beta$ for a specific total bandwidth. However, when the total bandwidth $B$ is sufficiently large (exceeding 40 MHz), the value of $\beta$ approaches 1 across all cases, as long as the mAP of the cloud server is greater than the mAP of the edge UAV. Conversely, when the total bandwidth $B$ is very small (not exceeding 1 MHz), the value of $\beta$ tends to approach 0 in all cases.

\begin{table}[!tpb]
\centering
\caption{Variables with different values of $N$}\label{Simulation6}
\begin{tabular}{|c|c|c|c|c|}
\hline
\makecell[c]{Number of frames \\per second $N$} & \textbf{5} & \textbf{10} & \textbf{15} & \textbf{20}\\
\hline
\makecell[c]{Uplink bandwidth \\$B_u$ (MHz)} & 10 & 8.55 & 6.57 & 6.44\\
\hline
\makecell[c]{Downlink bandwidth \\$B_d$ (MHz)} & 0 & 1.45 & 3.43 & 3.56\\
\hline
\makecell[c]{Task allocation ratio $\beta$ }& 1 & 0.385 & 0.193 & 0.142\\
\hline
\makecell[c]{Model update \\overhead $M$ (Mbps)} & $M_{min}$ & $7.25$ & $M_{max}$ & $M_{max}$\\
\hline
\makecell[c]{Average quantization bits\\ for residual mapping \\data $\bar{\bm{b}}$ (bit/pixel)} & 0.514 & 0.5662 & 0.5786 & 0.5782\\
\hline
\end{tabular}
\end{table}

In Table~\ref{Simulation6}, we evaluate the value of a few key variables corresponding to different number of frames generated per second. With $N=5$, the uplink bandwidth is the dominant factor, with all the tasks allocated to the cloud model. As the number of frames generated per second increases, the bandwidth allocated to downlink transmission grows, accommodating a rising proportion of classification tasks allocated to the edge model. That also fits the trend of variable $\beta$, which decreases with the increment of $N$. The overhead of model updates experiences a rapid increase from the minimum to the maximum value with the increment of $N$, aligning with a higher value of $mAP_S$. The value of the average quantization bits for residual mapping data does not change significantly with the increment of $N$, indicating that $mAP_L$ remains stable across different values of $N$ in the integrated air-ground edge-cloud model evolution framework.

\section{Conclusions}\label{Conclusions}
This paper has introduced a new integrated air-ground edge-cloud model evolution framework that enables concurrent edge model and cloud model data analysis, together with the ability to update a UAV edge model assisted by a ground cloud server. We have derived a closed-form expression for the lower bound of the mAP of the proposed framework, and have solved the mAP maximization problem by joint task allocation, transmission resource allocation, transmission data quantization, and edge model update design. Simulation results have underscored the superior mAP performance of the proposed framework across various communication bandwidths and data sizes, outperforming both a cloud model framework and an edge model framework. A few conclusions are summarised as follows.
\begin{enumerate}
\item The performance gain of the proposed framework stems from dynamically adjusting the mAP of the edge model and the cloud model via model evolution and data uploading, so as to maximize the overall mAP of the framework.
\item The edge model handles the majority of tasks with small communication bandwidth and large data size, where most of the bandwidth is allocated to small model updating.
\item The cloud model handles the majority of tasks with large communication bandwidth and small data size, with most of the bandwidth allocated to residual mapping data uploading.
\end{enumerate}

\begin{appendices}
\section{Proof of Theorem 1}
In adherence to the mAP definition, the mAP value corresponds to the area under the precision-recall curve (PRC), which is derived from a collection of paired precision-recall values across varying IoU thresholds. The precision is calculated as the ratio of true positive (TP) samples to the sum of TP and false positive (FP) samples, while the recall is determined by the ratio of TP samples to the sum of TP and false negative (FN) samples.

\begin{figure}[!tpb]
\centering
\includegraphics[width=2.4in]{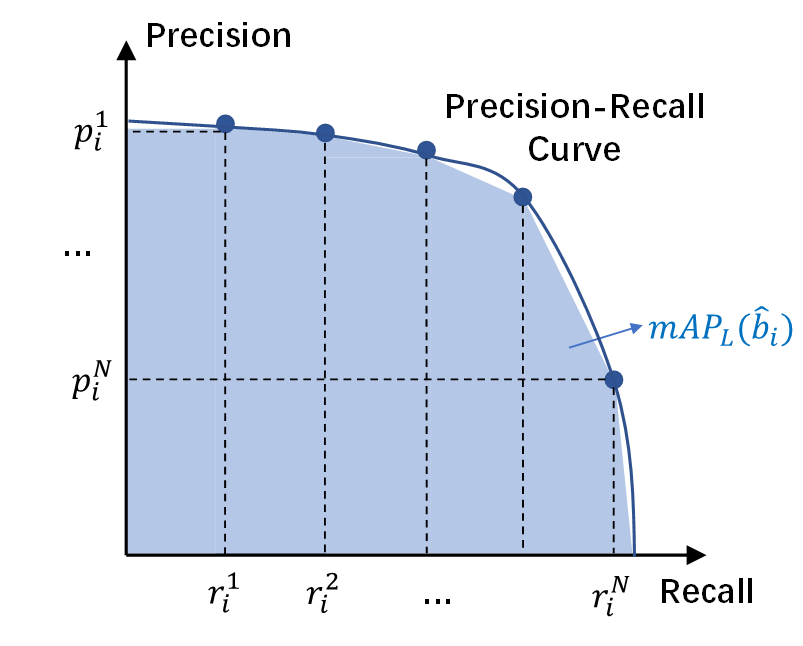}
\caption{Illustration for mAP and PRC.}
\label{AppendixAFig}
\end{figure}

We assume that frame $i$ and frame $j$ have different quantization bits for their residual mapping data transmission, denoted by $\hat{b}_i$ and $\hat{b}_j$, respectively. The mAP of the classification tasks at the cloud server with $\hat{b}_i$ and $\hat{b}_j$ are given as $mAP_L(\hat{b}_i)$ and $mAP_L(\hat{b}_j)$. Take frame $i$ as an example, the value of $mAP_L(\hat{b}_i)$ can be approximated as follows:
\begin{equation}\label{AppendixAmAPEqu}
mAP_L(\hat{b}_i)\approx \sum_{k=1}^K (r_i^k-r_i^{k-1})(p_i^k+p_i^{k-1})/2,
\end{equation}
where $r_i^k$ and $p_i^k$ are the recall and precision with the $k$th IoU threshold, as shown in Fig.~\ref{AppendixAFig}.

According to the definition of mAP, $r_i^k$ and $p_i^k$ can be expressed as $r_i^k=\frac{TP_i^k}{TP_i^k+FP_i^k}$, and $p_i^k=\frac{TP_i^k}{TP_i^k+FN_i^k}$, respectively, where $TP_i^k$ is the number of true positive samples, $FP_i^k$ is the number of false positive samples, and $FN_i^k$ is the number of false negative samples. With $x_i^k=\frac{FP_i^k}{TP_i^k}$ and $y_i^k=\frac{FN_i^k}{TP_i^k}$, we have $r_i^k=\frac{1}{1+x_i^k}$ and $p_i^k=\frac{1}{1+y_i^k}$. Similarly, for frame $j$, variables $r_j^k$ and $p_j^k$ can be expressed as $r_j^k=\frac{1}{1+x_j^k}$ and $p_j^k=\frac{1}{1+y_j^k}$, respectively.

With joint consideration to the mAP performance of frame $i$ and frame $j$, the recall value with the $k$th IoU threshold is $r^k=\frac{2}{2+x_i^k+x_j^k}$. We then compare the value of $r^k$ and the average value of the recall values of frame $i$ and frame $j$ as follows:
\begin{equation}\label{AppendixAEq1}
\begin{split}
&\frac{r_i^k+r_j^k}{2}-r^k=\frac{1}{1+x_i^k}+\frac{1}{1+x_j^k}-\frac{2}{2+x_i^k+x_j^k}\\
&=\frac{((1+x_i^k)+(1+x_j^k))^2-2(1+x_i^k)(1+x_j^k)}{(1+x_i^k)(1+x_j^k)(2+x_i^k+x_j^k)}\\
&\geq\frac{(x_i^k-x_j^k)^2}{(1+x_i^k)(1+x_j^k)(2+x_i^k+x_j^k)}\geq 0.
\end{split}
\end{equation}
The relation in~(\ref{AppendixAEq1}) shows that the recall value of the two frames is less than the average of that of the two frames separately. Similarly, we can also obtain the relation
\begin{equation}\label{AppendixAEq2}
\frac{p_i^k+p_j^k}{2}-p^k\geq 0.
\end{equation}
When substituting~(\ref{AppendixAEq1}) and (\ref{AppendixAEq2}) into~(\ref{AppendixAmAPEqu}), we conclude that the mAP of the two frames is less than the average of that of the two frames separately. Moreover, \emph{Assumption 1} shows that the mAP is a concave function with respect to the quantization bits of residual mapping data. We then have the following relation
\begin{equation}\label{ConRel}
mAP(\hat{b}_i,\hat{b}_j)\leq \frac{mAP(\hat{b}_i)+mAP(\hat{b}_j)}{2}< mAP(\frac{\hat{b}_i+\hat{b}_j}{2}),
\end{equation}
where $mAP(\hat{b}_i,\hat{b}_j)$ is the joint mAP performance of frame $i$ and frame $j$. The inequality in~(\ref{ConRel}) shows that the mAP of two frames with different quantization bits of residual mapping is less than that of two frames with the same quantization bits. As a result, \emph{Theorem 1} holds.

\section{Proof of Theorem 2}
As proved in Appendix A, the quantization bits of the residual mapping data of all frames in the set $\Phi$ are the same. We compare the following two cases that satisfy \emph{Theorem 1}.
\begin{enumerate}
\item \emph{Case 1:} The residual mapping data of a ratio of $\rho$ $(0<\rho<1)$ frames are transmitted to the cloud server with the quantization bits of $\hat{b}$, and the residual mapping data of the other $1-\rho$ frames are not transmitted to the cloud server.
\item \emph{Case 2:} The residual mapping data of all frames are transmitted to the cloud server with the quantization bits of $\rho\hat{b}$.
\end{enumerate}
We denote the mAP of all the frames in $\Psi$ in case 1 by $mAP(0_{|1-\rho},\hat{b}_{|\rho})$. According to the result derived in \emph{Theorem 1}, when \emph{Assumption 2} is satisfied, the mAP of the two cases satisfies
\begin{equation}\label{AppendixBEq1}
mAP(0_{|1-\rho},\hat{b}_{|\rho})\leq (1-\rho)mAP(0)+\rho mAP(\hat{b})< mAP(\rho\hat{b}).
\end{equation}
The inequality in~(\ref{AppendixBEq1}) shows that the mAP of case 2 is larger than that of case 1. In other word, the mAP of the case of $\rho=1$ outperforms that of the case of $0<\rho<1$. Therefore, to maximize the mAP at the cloud server, the residual mapping data of all frames in set $\Phi$ are sent to the cloud server with the same quantization bits, i.e., $\rho=1$, and \emph{Theorem 2} holds.

\section{Proof of Theorem 3}
As discussed in Appendix A, the mAP is a function of the precision-recall pairs of different IoU thresholds. To study the relation between $mAP$ and the precision-recall pairs of the two models, we first analyse the expression of the joint precision and recall values of the integrated air-ground edge-cloud model evolution framework. Denote the precision and recall values of the integrated air-ground edge-cloud model evolution framework for the $k$th IoU threshold by $r^k=\frac{TP^k}{TP^k+FP^k}$ and $p^k=\frac{TP^k}{TP^k+FN^k}$, respectively.

Take the precision value as an example, as analysed in Appendix A, the precision value of the cloud model for the $k$th IoU threshold can be expressed as
\begin{equation}\label{AppendixCEq1}
p_L^k=\frac{1}{1+y_L^k},
\end{equation}
with $y_L^k=\frac{FN_L^k}{TP_L^k}$, and the precision value of the edge model for the $k$th IoU threshold can be expressed as
\begin{equation}\label{AppendixCEq2}
p_S^k=\frac{1}{1+y_S^k},
\end{equation}
with $y_S^k=\frac{FN_S^k}{TP_S^k}$. Equations~(\ref{AppendixCEq1}) and~(\ref{AppendixCEq2}) shows that each TP sample at the cloud model is accompanied by $y_L^k$ FN samples, and each TP sample at the edge model is accompanied by $y_S^k$ FN samples. With the number of samples at the two models being $\frac{\beta}{1-\beta}$, the average precision value can be expressed as
\begin{equation}\label{AppendixCEq3}
\begin{split}
p^k&=\frac{TP^k}{TP^k+FN^k}=\frac{\beta+(1-\beta)}{\beta(1+y_L^k)+(1-\beta)(1+y_S^k)}\\
&=\frac{1}{1+\beta y_L^k+(1-\beta)y_S^k}\\
&=\frac{1}{1+\beta (\frac{1}{p_L^k}-1)+(1-\beta)(\frac{1}{p_S^k}-1)}\\
&=\frac{1}{\frac{\beta}{p_L^k}+\frac{1-\beta}{p_S^k}}.
\end{split}
\end{equation}
Similarly, the recall value has a relation of
\begin{equation}\label{AppendixCEq4}
r^k=\frac{1}{\frac{\beta}{r_L^k}+\frac{1-\beta}{r_S^k}}.
\end{equation}
Equation~(\ref{Theorem3}) can be obtained by substituting~(\ref{AppendixCEq3}) and~(\ref{AppendixCEq4}) into~(\ref{AppendixAmAPEqu}), and \emph{Lemma 1} is proved.

\section{Proof of Theorem 4}
In equation~(\ref{Theorem3}), the mAP is a linear summation of a function of the precision-recall pairs for different IoU thresholds. We can study the relation among $mAP$, $mAP_L$ and $mAP_S$ by analysing the precision-recall pair of a specific IoU threshold, and the property still holds with linear transformations. Define a variable $\zeta^k=p^k r^k$ which is only related to the precision-recall pair of a specific IoU threshold. Correspondingly, the variable of the cloud model at the server and the edge model at the UAV are denoted by $\zeta_L^k=p_L^k r_L^k$ and $\zeta_S^k=p_S^k r_S^k$, respectively. According to \emph{Lemma 1}, $\zeta^k$ can be expressed as
\begin{equation}\label{AppendixDEq1}
\begin{split}
&\zeta^k=p^k r^k=\frac{1}{\frac{\beta}{p_L^k}+\frac{1-\beta}{p_S^k}} \cdot \frac{1}{\frac{\beta}{r_L^k}+\frac{1-\beta}{r_S^k}}\\
&=\frac{p_L^k p_S^k}{\beta p_S^k+(1-\beta)p_L^k} \cdot \frac{r_L^k r_S^k}{\beta r_S^k+(1-\beta)r_L^k}\\
&=\frac{p_L^k p_S^k r_L^k r_S^k}{\beta^2 p_S^k r_S^k + (1-\beta)^2 p_L^k r_L^k + \beta(1-\beta) (p_L^k r_S^k + p_S^k r_L^k)}\\
&=\frac{\zeta_L^k \zeta_S^k}{\beta^2 \zeta_S^k + (1-\beta)^2 \zeta_L^k + \beta(1-\beta) (\zeta_L^k + \zeta_S^k - (p_L^k r_L^k - p_S^k r_S^k))}\\
&=\frac{\zeta_L^k \zeta_S^k}{(1-\beta)\zeta_L^k+\beta \zeta_S^k-\beta(1-\beta)\Delta},
\end{split}
\end{equation}
where $\Delta=(p_L^k - p_S^k)\cdot (r_L^k - r_S^k)$ shows the inference capacity difference between the cloud model and the edge model. Considering that the inference capacity of the cloud model is better than the edge model, it is reasonable to have $p_L^k - p_S^k>0$ and $r_L^k - r_S^k>0$. Therefore, relation $\Delta>0$ holds, and thus we have
\begin{equation}\label{AppendixDEq2}
\zeta^k \geq \frac{\zeta_L^k \zeta_S^k}{(1-\beta)\zeta_L^k+\beta \zeta_S^k}.
\end{equation}
Since $mAP$ is a linear combination of a series of $\zeta^k$, the relation in~(\ref{AppendixDEq2}) also holds for the $mAP$, i.e.,
\begin{equation}\label{AppendixDEq3}
mAP \geq \frac{mAP_L mAP_S}{(1-\beta)mAP_L+\beta mAP_S},
\end{equation}
and \emph{Theorem 4} is proved.

\section{Proof of Theorem 7}
When the mAP changing rate of options 1 and 2 are equal, the following equation holds with \emph{Theorem 2},
\begin{equation}\label{mAPEqual1}
\frac{\emph{d}(mAP)}{\emph{d}(M/S_d)}=\frac{\emph{d}(mAP)}{\emph{d}(|\Psi|\hat{b}_i/S_u)}.
\end{equation}
When focusing on the case that satisfies \emph{Theorem 5}, we substitute (\ref{mAPApproximate}) into (\ref{mAPEqual1}), and have
\begin{equation}\label{mAPEqual2}
\frac{(1-\beta)mAP_L^2}{S_u}\cdot\frac{\emph{d} (h(M))}{\emph{d} M}=\frac{\beta mAP_S^2}{|\Psi|S_d}\cdot\frac{\partial g(\rho, \hat{b}_i)}{\partial \hat{b}_i}.
\end{equation}

Similarly, when the mAP changing rates of options 1 and 3 are equal, the following equation holds with \emph{Theorem 2},
\begin{equation}\label{mAPEqual3}
\frac{\emph{d}(mAP)}{\emph{d}(M/S_d)}=\frac{\emph{d}(mAP)}{\emph{d}((\bar{F}+\hat{b}_i)/S_u)}.
\end{equation}
When focusing on the case that satisfies \emph{Theorem 5}, we substitute (\ref{mAPApproximate}) into (\ref{mAPEqual1}), and have
\begin{equation}\label{mAPEqual4}
\frac{(1-\beta)mAP_L}{S_u}\cdot\frac{\emph{d} (h(M))}{\emph{d} M}=\frac{mAP_L\cdot mAP_S - mAP_S^2}{N(\bar{F}+\hat{b}_i)S_d}.
\end{equation}
Variable $\beta$ can be eliminated by combining~(\ref{mAPEqual2}) and~(\ref{mAPEqual4}). The relation between $mAP_L$ and $mAP_S$ can be expressed as~(\ref{Theorem5Equation}) in \emph{Theorem 7}.

\end{appendices}

\end{document}